\documentclass[journal]{IEEEtran}
\ifCLASSINFOpdf
\else
\fi
\usepackage{amsmath,amssymb,amsfonts}
\usepackage{cite}

\usepackage{algorithmic}
\usepackage{stfloats}
\usepackage{graphicx}
\usepackage{subfigure}
\usepackage{textcomp}
\usepackage{xcolor}
\usepackage{pifont}
\usepackage{amsthm}
\newtheorem{theorem}{Theorem}

\usepackage{mathrsfs}
\usepackage{cases}
\usepackage{comment}
\usepackage{empheq} 
\usepackage{caption}
\usepackage{bm}
\allowdisplaybreaks
\usepackage{xcolor}
\usepackage{url}  
\usepackage{graphicx}  
\usepackage[ruled,vlined,linesnumbered]{algorithm2e}
\usepackage{setspace}
\usepackage{floatpag} 
\usepackage{float}
\floatpagestyle{empty}

\usepackage{listings}
\usepackage{fancyhdr} 
\usepackage{url}

\begin{document}
%
\title{{{}{}Secure Beamforming for Multiple Intelligent Reflecting Surfaces Aided mmWave Systems}}
%
%
%

\author{Yue Xiu,~Jun Zhao,~\IEEEmembership{Member,~IEEE},
~Chau Yuen,~\IEEEmembership{Fellow,~IEEE},~Zhongpei Zhang,~\IEEEmembership{Member,~IEEE},~Guan Gui,~\IEEEmembership{Senior Member,~IEEE}
\thanks{Yue Xiu, and Zhongpei Zhang are with University of Electronic Science and Technology of China, Chengdu, China (E-mail: xiuyue@std.uestc.edu.cn).
Jun Zhao is with Nanyang Technological University, Singapore (E-mail: junzhao@ntu.edu.sg).
Chau Yuen is with
Singapore University of Technology and Design (SUTD), Singapore (E-mail: yuenchau@sutd.edu.sg).
Guan Gui is with Nanjing University of Posts and Telecommunications, Nanjing, China (E-mail: guiguan@njupt.edu.cn).

This work was supported by 1) Guangdong province Key Project of science and Technology(2018B010115001), 2) Major Project of the Ministry of Industry and Information Technology of China under Grant (TC190A3WZ-2), Nanyang Technological University (NTU) Startup Grant, 3) Alibaba-NTU Singapore Joint Research Institute (JRI), 4) Singapore National Research Foundation (NRF) under its Strategic Capability Research Centres Funding Initiative: Strategic Centre for Research in Privacy-Preserving Technologies \& Systems (SCRIPTS).
}}

\maketitle

	 \thispagestyle{fancy}
\pagestyle{fancy}
\lhead{This paper appears in \textbf{IEEE Communications Letters}. \hfill \thepage \\ Please feel free to contact us for questions or remarks. \hfill \url{https://doi.org/10.1109/LCOMM.2020.3028135/} }
\cfoot{}
\renewcommand{\headrulewidth}{0.4pt}
\renewcommand{\footrulewidth}{0pt}

\begin{abstract}
In this letter, {{}{}secure beamforming in a multiple intelligent reflecting surfaces (IRSs)-aided millimeter-wave (mmWave) system is investigated. In this system, the secrecy rate is maximized by controlling the on-off status of each IRS as well as optimizing the phase shift matrix of the IRSs}. This problem is posed as a joint optimization problem of transmit beamforming and IRS control, whose goal is to maximize the secrecy rate under the total transmission power and unit-modulus constraints.  The problem is difficult to solve optimally due to the nonconvexity of constraint conditions and coupled variables. To deal with this problem, we propose an alternating optimization (AO)-based algorithm based on successive convex approximation (SCA) and manifold optimization (MO) technologies. Numerical simulations show that the proposed AO-based algorithm can effectively improve the secrecy rate and outperforms traditional single IRS-aided scheme.
\end{abstract}

\begin{IEEEkeywords}
Multiple intelligent reflecting surfaces, millime ter-wave, secrecy rate, alternating optimization.
\end{IEEEkeywords}

%
\IEEEpeerreviewmaketitle

\section{Introduction}
Millimeter-wave (mmWave) technologies have been a key technology for 5G communications, due to its abundant spectrum and high data rates \cite{xiao2017millimeter,xiu2020irs}. However, due to the high propagation loss, mmWave signals are easily blocked by obstacles. Recently, intelligent reflecting surface (IRS) becomes a promising technology for solving these problems \cite{wu2018intelligent,huang2019reconfigurable,hong2020artificial,li2020weighted,lu2020robust,zhou2020framework}. Specifically, by adjusting the 
phase shifts of the IRS, the transmission signals are able to be strengthened. Moreover, since IRS can significantly improve the beamforming gain, the IRS can extend the coverage of mmWave communication systems as well \cite{yildirim2019propagation,basar2019wireless}.  

On the other hand, the secrecy rate optimization problem has been intensively investigated in recent years. Because phase shifts of the IRS can configure the wireless channels, it can greatly improve the secrecy rate \cite{qiao2020secure,zhou2020robust}. Specifically, {{}{}an IRS-aided mmWave secure single-user multiple-input single-output (MISO) system was investigated in \cite{qiao2020secure}. In \cite{qiao2020secure}, the IRS's phase shifts were adaptively adjusted to strengthen the received signal at the user but suppressed the eavesdropper. In \cite{zhou2020robust}, a robust secure based on block coordinate descent (BCD) algorithm was proposed for maximizing the secrecy rate in an IRS-aided mmWave MISO communication systems}.

However, only a single IRS was considered in the above-mentioned works. Due to the limited coverage of a single IRS, one IRS can not be satisfied with the users’ high quality service requirements. Deploying a number of low-cost power-efficient IRSs in future wireless systems can cooperatively further improve the performance of the systems [13]. 
This motivates us to investigate a multiple IRSs-aided mmWave system with switches. 
Specifically, assuming that the channels of
user and eavesdropper is perfectly known at an access point (AP). The secrecy rate of the system is maximized by jointly optimizing the phase shifts of all IRSs, the
transmit beamforming of the transmitter, and the IRS on-off status vector. Because the formulated problem is non-convex, we propose an alternating optimization (AO)-based algorithm to solve it. 


\section{system model}
\begin{figure}[!t]
\centering
\includegraphics[height=1.2in,width=2.2in]{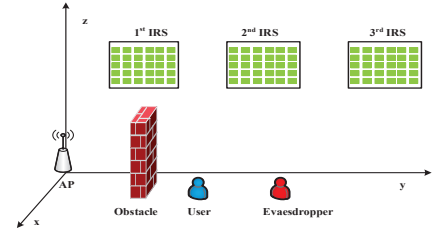}
\caption{System model for mmWave communication system with multiple IRSs.\vspace{-20pt}}
\label{fig:1}
\end{figure}
As shown in Fig.~\ref{fig:1}, we consider a mmWave downlink with multiple IRSs, which consists of one AP, a set $\mathcal{L}$ of $L$ IRSs, one user, and one eavesdropper. The AP is equipped with $N_{t}$ antennas. Each IRS, $l\in\mathcal{L}$, has $N_{r}$ reflecting elements. The user and eavesdropper are equipped with a single antenna, respectively.  {{}{}In this paper, we assume that the direct link
between the AP and the user or the eavesdropper is blocked by obstacles, which usually occurs when the direct link is blocked due to long-distance path loss or obstacles\cite{xiao2017millimeter}}.

The  AP-to-$l$th IRS  mmWave channel, the $l$th IRS-to-user mmWave channel, and  the $l$th IRS-to-eavesdropper mmWave channel are denoted as $\boldsymbol{G}_{l}\in\mathbb{C}^{N_{r}\times N_{t}}$, $\boldsymbol{h}_{l}\in\mathbb{C}^{N_{r}\times 1}$, $\boldsymbol{g}_{l}\in\mathbb{C}^{N_{r}\times 1}$, respectively. The received signal at the $l$th IRS is expressed as $\boldsymbol{r}_{l}=\boldsymbol{G}_{l}\boldsymbol{w}s$,
{{}{}where $\boldsymbol{r}_{l}\in\mathbb{C}^{N_{r}\times 1}$. $s\in\mathbb{C}^{1\times 1}$ and $\boldsymbol{w}\in\mathbb{C}^{N_{t}\times 1}$ denote the transmit data and the corresponding beamforming vector at the AP with $\mathbb{E}[ss^{H}]=1$. Then the $l$th IRS reflects it with a phase shift matrix $\boldsymbol{\Theta}_{l}=\mathrm{diag}(\boldsymbol{\theta}_{l})\in\mathbb{C}^{N_{r}\times N_{r}}$, where $\boldsymbol{\theta}_{l}=[\theta_{l,1},\cdots,\theta_{l,N_{r}}]^{T}\in\mathbb{C}^{N_{r}\times 1}$ and $\theta_{l,j}=e^{j\phi_{j}}, j=1,\cdots,N_{r}$ with $\phi_{j}$ being the reflection phase shift.} The received signals at the user and the eavesdropper are denoted as
\begin{align}
&{}{y=\sum\nolimits_{l=1}^{L}x_{l}\boldsymbol{h}_{l}^{H}\boldsymbol{\Theta}_{l}\boldsymbol{G}_{l}\boldsymbol{w}s+n},\label{4-2}\\
&{}{y_{e}=\sum\nolimits_{l=1}^{L}x_{l}\boldsymbol{g}_{l}^{H}\boldsymbol{\Theta}_{l}\boldsymbol{G}_{l}\boldsymbol{w}s+n_{e}}\label{4-3},
\end{align}
where $x_{l}\in\{0,1\}$ is a binary variable.  When $x_{l}=1$, the $l$th IRS is active. When $x_{l}=0$, the $l$th IRS does not work and consume no power. $n\sim\mathcal{CN}(0,\sigma^{2})$ and $n_{e}\sim\mathcal{CN}(0,\sigma_{e}^{2})$ denote additive Gaussian noise of the user and the eavesdropper, respectively.

According to (\ref{4-2}) and (\ref{4-3}), the achievable rate at the user is expressed as
\begin{eqnarray}
{}{I=\log_{2}(1+\frac{1}{\sigma^{2}}\|\sum\nolimits_{l=1}^{L}x_{l}\boldsymbol{h}_{l}^{H}\boldsymbol{\Theta}_{l}\boldsymbol{G}_{l}\boldsymbol{w}\|^{2})}.\label{4-4}
\end{eqnarray}
The achievable rate at the eavesdropper is expressed as 
\begin{eqnarray}
{}{I_{e}=\log_{2}(1+\frac{1}{\sigma_{e}^{2}}\|\sum\nolimits_{l=1}^{L}x_{l}\boldsymbol{g}_{l}^{H}\boldsymbol{\Theta}_{l}\boldsymbol{G}_{l}\boldsymbol{w}\|^{2})}.\label{4-5}
\end{eqnarray}
Therefore, the secrecy rate $I_{s}$ can be denoted as
\begin{equation}
{}{I_{s}=[I-I_{e}]^{+}},\label{4-6}
\end{equation}
where $[x]^{+}=\max(0,x)$.
The transmit power constraint is
\begin{eqnarray}
{}{\mathrm{tr}\left(\boldsymbol{w}\boldsymbol{w}^{H}\right)\leq P},\label{4-7}
\end{eqnarray}
where $P$ is the maximum transmit power of the AP. 
Based on (\ref{4-4})-(\ref{4-7}), the secure beamforming optimization problem for multiple IRSs-aided mmWave system with power constraint is formulated as
\begin{subequations}
\begin{align}
{}{\max\limits_{\boldsymbol{w},\boldsymbol{\theta},\boldsymbol{x}}}~&{}{I_{s},}\label{4-8a}\\
{}{\mbox{s.t.}}~
&{}{\mathrm{tr}\left(\boldsymbol{w}\boldsymbol{w}^{H}\right)\leq P,}\label{4-8b}&\\
&{}{|\theta_{l,j}|=1,~\forall l\in\mathcal{L},j=1,\cdots,N_{r}.}&\label{4-8c}\\
&{}{x_{l}\in\{0,1\}, \forall l\in\mathcal{L},}&\label{4-8d}
\end{align}\label{4-8}%
\end{subequations}
where $\boldsymbol{x}=[x_{1},\cdots,x_{L}]^{T}$.
Problem (\ref{4-8}) is highly non-convex because of the non-convexity of the objective function and constraints. In the following section, we propose one iterative algorithm to obtain suboptimal solutions.


\section{AO-based Algorithm for Problem}
\subsection{Transmit Beamforming Optimization}
In this subsection, we first fix the variables $\boldsymbol{\theta}$ and $\boldsymbol{x}$, problem (\ref{4-8}) is rewritten as
\begin{subequations}
\begin{align}
{}{\max\limits_{\boldsymbol{w}}}~&{}{I_{s}},\label{4-9a}\\
{}{\mbox{s.t.}}~
&{}{\mathrm{tr}\left(\boldsymbol{w}\boldsymbol{w}^{H}\right)\leq P.}\label{4-9b}
\end{align}\label{4-9}%
\end{subequations}
Next, we introduce a new matrix $\boldsymbol{W}=\boldsymbol{w}\boldsymbol{w}^{H}$. Let $\boldsymbol{a}^{H}=\sum\nolimits_{l=1}^{L}x_{l}\boldsymbol{h}_{l}^{H}\boldsymbol{\Theta}_{l}\boldsymbol{G}_{l}$ and $\boldsymbol{b}^{H}=\sum\nolimits_{l=1}^{L}x_{l}\boldsymbol{g}_{l}^{H}\boldsymbol{\Theta}_{l}\boldsymbol{G}_{l}$. Thus, we have $\|\sum\nolimits_{l=1}^{L}x_{l}\boldsymbol{h}_{l}^{H}\boldsymbol{\Theta}_{l}\boldsymbol{G}_{l}\boldsymbol{w}\|^{2}=\mathrm{Tr}(\boldsymbol{W}\boldsymbol{A})$ and $\|\sum\nolimits_{l=1}^{L}x_{l}\boldsymbol{g}_{l}^{H}\boldsymbol{\Theta}_{l}\boldsymbol{G}_{l}\boldsymbol{w}\|^{2}=\mathrm{Tr}(\boldsymbol{W}\boldsymbol{B})$, where $\boldsymbol{A}=\boldsymbol{a}\boldsymbol{a}^{H}$ and $\boldsymbol{B}=\boldsymbol{b}\boldsymbol{b}^{H}$. It is not difficult to find that if $\boldsymbol{W}$ 
is regarded as an optimization variable, $\boldsymbol{W}$ needs to satisfy the rank one constraint, i.e., $\mathrm{rank}(\boldsymbol{W})=1$. Then, the semi-definite relaxation (SDR) technology is used to omit the constraint $\mathrm{rank}(\boldsymbol{W})=1$. Thus, (\ref{4-9}) can be rewritten as
\begin{subequations}
\begin{align}
{}{\max\limits_{\boldsymbol{w}}}~&{}{\log_{2}\left(\frac{\frac{1}{\sigma^{2}}\mathrm{Tr}(\boldsymbol{W}\boldsymbol{A})+1}{\frac{1}{\sigma^{2}_{e}}\mathrm{Tr}(\boldsymbol{W}\boldsymbol{B})+1}\right),}\label{4-10a}\\
{}{\mbox{s.t.}}~
&{}{\mathrm{tr}\left(\boldsymbol{W}\right)\leq P,}&\label{4-10b}\\
&{}{\boldsymbol{W}\succeq\boldsymbol{0},~i=1,\cdots,K.}\label{4-10c}
\end{align}\label{4-10}%
\end{subequations}
However, (\ref{4-10}) is still non-convex due to the non-convexity of (\ref{4-10a}) is non-convex. To handle the non-convex parts, we introduce the following auxiliary variables
\begin{align}
{}{e^{p}=1+\frac{1}{\sigma^{2}}\mathrm{Tr}(\boldsymbol{W}\boldsymbol{A})},~~~{}{e^{q}=1+\frac{1}{\sigma_{e}^{2}}\mathrm{Tr}(\boldsymbol{W}\boldsymbol{B})}.\label{4-12}
\end{align}
By substituting (\ref{4-12}) into (\ref{4-10}), we can reformulate the problem in (\ref{4-10}) as follows.
\begin{subequations}
\begin{align}
{}{\max\limits_{\boldsymbol{W},p,q}}~&{}{\log_{2}\left(e^{p-q}\right),}\label{4-13a}\\
\mbox{s.t.}~
&{}{1+\frac{1}{\sigma^{2}}\mathrm{Tr}(\boldsymbol{W}\boldsymbol{A})\geq e^{p},}&\label{4-13b}\\
&{}{1+\frac{1}{\sigma_{e}^{2}}\mathrm{Tr}(\boldsymbol{W}\boldsymbol{B})\leq e^{q},}&\label{4-13c}\\
&{}{\mathrm{tr}\left(\boldsymbol{W}\right)\leq P,\boldsymbol{W}\succeq\boldsymbol{0},}&\label{4-13d}\\
&{}{\mathrm{Tr}({\boldsymbol{W}\boldsymbol{A}})\geq 0,\mathrm{Tr}({\boldsymbol{W}\boldsymbol{B}})\geq 0.}&\label{4-13e}
\end{align}\label{4-13}%
\end{subequations}

According to the properties of the logarithmic function, the objective function in (\ref{4-13a}) can be expressed as
\begin{align}
{}{\log_{2}\left(e^{p-q}\right)=(p-q)\log_{2}(e).}\label{4-14}
\end{align}
Thus, (\ref{4-14}) is linear and convex. We replace the equalities in (\ref{4-12}) with the inequalities in (\ref{4-13b}) and (\ref{4-13c}). It is not difficult to find that because of the monotonicity of the objective function, inequalities (\ref{4-13b}) to (\ref{4-13c}) would hold with equalities at optimal points. 

It can be observed that (\ref{4-13c}) is non-convex, to deal with the non-convex constraint, we consider the successive convex approximation (SCA) algorithm. The first-order Taylor expansion of $e^{q}$ at $\bar{q}$ is given by
\begin{eqnarray}
{}{e^{\bar{q}}+e^{\bar{q}}(q-\bar{q}).}\label{4-15}
\end{eqnarray}
Thus, the constraint condition in (\ref{4-13c}) can be rewritten as
\begin{eqnarray}
{}{1+\frac{1}{\sigma_{e}^{2}}\mathrm{Tr}(\boldsymbol{W}\boldsymbol{B})\leq e^{\bar{q}}+e^{\bar{q}}(q-\bar{q}).}\label{4-16}
\end{eqnarray}
By replacing (11c) with (\ref{4-16}), the following problem can be obtained. 
\begin{subequations}
\begin{align}
{}{\max\limits_{\boldsymbol{W},p,q}}~&{}{\log_{2}\left(e^{p-q}\right),}\label{4-17a}\\
{}{\mbox{s.t.}}~
&{}{1+\frac{1}{\sigma_{e}^{2}}\mathrm{Tr}(\boldsymbol{W}\boldsymbol{B})\leq e^{\bar{q}}+e^{\bar{q}}(q-\bar{q}),}&\label{4-17c}\\
&{}{\text{(\ref{4-13b})},\text{(\ref{4-13d})},\text{(\ref{4-13e})}.}&\label{4-17d}
\end{align}\label{4-17}%
\end{subequations}
Now, problem (\ref{4-17}) is convex, CVX\cite{boyd2004convex} can be used to solve this problem efficiently. In the $t$th iteration, the convex approximate problem is expressed as
\begin{subequations}
\begin{align}
{}{\max\limits_{\boldsymbol{W},p,q}}~&{}{\log_{2}\left(e^{p-q}\right),}\label{4-18a}\\
{}{\mbox{s.t.}}~
&{}{1+\frac{1}{\sigma_{e}^{2}}\mathrm{Tr}(\boldsymbol{W}\boldsymbol{B})\leq e^{\bar{q}^{t}}+e^{\bar{q}^{t}}(q-\bar{q}^{t}),}&\label{4-18c}\\
&{}{\text{(\ref{4-13b})},\text{(\ref{4-13d})},\text{(\ref{4-13e})}.}&\label{4-18e}
\end{align}\label{4-18}%
\end{subequations}
When update $\bar{q}$, let $\bar{q}^{t+1}=q^{t}$. It should be noticed that to get the initial value $\bar{q}^{1}$, we first generate $\boldsymbol{w}^{0}$ randomly, and compute $\boldsymbol{W}^{0}=\boldsymbol{w}^{0}(\boldsymbol{w}^{0})^{H}$ . The SCA-based algorithm is summarized in \textbf{Algorithm}~\ref{algo-1}.
\begin{algorithm}[!htbp]\small
\caption{SCA-based Algorithm for Problem (\ref{4-10})}
   \begin{algorithmic}[1]
   \REQUIRE $t=0$,  
given $\boldsymbol{w}^{0}$ that is satisfy conditions, calculate $q^{0}$ based on (\ref{4-12}) and let $\bar{q}^{1}=q^{0}$.\\ 
   \STATE \textbf{repeat}
     \STATE Solve problem in (\ref{4-18}) to obtain the optimal solution $\boldsymbol{W}^{t}$ and $q^{t}$.\\
Update $\bar{q}^{t+1}=q^{t}$.\\
    \STATE Update $\bm{W}^{t+1}$ by solving (16);\\
    \STATE Set $t=t+1$.\\
     \STATE \textbf{until} the stopping criterion is met.\\
     \ENSURE {{}{}$\boldsymbol{w}$ is obtained by decomposition of $\boldsymbol{W}$ when $\mathrm{rank}(\boldsymbol{W})=1$; Otherwise, we use Gaussian Random to get $\boldsymbol{w}$.}\\
   \end{algorithmic}\label{algo-1}
 \end{algorithm}\vspace{-20pt}

\subsection{Phase Shift and IRS On-Off Optimization}
{{}{}Given the transmit beamforming variable $\boldsymbol{W}$ obtained in the previous section and $\{\boldsymbol{\Theta}_{l}\}$, problem (\ref{4-8}) becomes}\par
\begin{subequations}
\begin{align}
{}{\max\limits_{\boldsymbol{x}}}~&{}{\frac{1+\frac{1}{\sigma^{2}}\left\|\sum\nolimits_{l=1}^{L}x_{l}\boldsymbol{h}_{l}^{H}\boldsymbol{\Theta}_{l}\boldsymbol{G}_{l}\boldsymbol{w}\right\|^{2}}{1+\frac{1}{\sigma_{e}^{2}}\left\|\sum\nolimits_{l=1}^{L}x_{l}\boldsymbol{g}_{l}^{H}\boldsymbol{\Theta}_{l}\boldsymbol{G}_{l}\boldsymbol{w}\right\|^{2}},}\label{4-19a}\\
\mbox{s.t.}~
&\text{(\ref{4-8d}).}\label{4-19b}
\end{align}\label{4-19}%
\end{subequations}
There are two difficulties in solving problem (\ref{4-19}). The first one is that objective function (\ref{4-19a}) is non-convex. The second one is that constraint (\ref{4-8d}) is non-convex. To deal with the first difficulty, we rewrite $\left\|\sum\nolimits_{l=1}^{L}x_{l}\boldsymbol{h}_{l}^{H}\boldsymbol{\Theta}_{l}\boldsymbol{G}_{l}\boldsymbol{w}\right\|^{2}$ and $\left\|\sum\nolimits_{l=1}^{L}x_{l}\boldsymbol{g}_{l}^{H}\boldsymbol{\Theta}_{l}\boldsymbol{G}_{l}\boldsymbol{w}\right\|^{2}$ as
\begin{eqnarray}
&{}{\|\sum\nolimits_{l=1}^{L}x_{l}\boldsymbol{h}_{l}^{H}\boldsymbol{\Theta}_{l}\boldsymbol{G}_{l}\boldsymbol{w}\|^{2}=\sum\nolimits_{l=1}^{L}C_{l}x_{l}+\sum\nolimits_{l=2}^{L}\sum\nolimits_{m=1}^{l-1}}&\nonumber\\
&{}{C_{lm}x_{l}x_{m}}~~~~~~~~~~~~~~~~~~~~~~~~~~~~~~~~~~~~~~~~~~~~~~~~~~~&\label{4-20}\\
&{}{\|\sum\nolimits_{l=1}^{L}x_{l}\boldsymbol{g}_{l}^{H}\boldsymbol{\Theta}_{l}\boldsymbol{G}_{l}\boldsymbol{w}\|^{2}=\sum\nolimits_{l=1}^{L}D_{l}x_{l}+\sum\nolimits_{l=2}^{L}\sum\nolimits_{m=1}^{l-1}}&\nonumber\\
&{}{D_{lm}x_{l}x_{m},}~~~~~~~~~~~~~~~~~~~~~~~~~~~~~~~~~~~~~~~~~~~~~~~~~~&\label{4-21}
\end{eqnarray}
where $C_{lm}=\boldsymbol{h}_{l}^{H}\boldsymbol{\Theta}_{l}\boldsymbol{G}_{l}\boldsymbol{w}\boldsymbol{w}_{m}^{H}\boldsymbol{G}_{m}^{H}\boldsymbol{\Theta}_{m}^{H}\boldsymbol{h}_{m}$ and $D_{lm}=\boldsymbol{g}_{l}^{H}\boldsymbol{\Theta}_{l}\boldsymbol{G}_{l}\boldsymbol{w}\boldsymbol{w}_{m}^{H}\boldsymbol{G}_{m}^{H}\boldsymbol{\Theta}_{m}^{H}\boldsymbol{g}_{m}$, $C_{l}=\boldsymbol{h}_{l}^{H}\boldsymbol{\Theta}_{l}\boldsymbol{G}_{l}\boldsymbol{w}\boldsymbol{w}_{l}^{H}\boldsymbol{G}_{l}^{H}\boldsymbol{\Theta}_{l}^{H}\boldsymbol{h}_{l}$ and $D_{l}=\boldsymbol{g}_{l}^{H}\boldsymbol{\Theta}_{l}\boldsymbol{G}_{l}\boldsymbol{w}\boldsymbol{w}_{l}^{H}\boldsymbol{G}_{l}^{H}\boldsymbol{\Theta}_{l}^{H}\boldsymbol{g}_{l}$. 
We use the parametric approach in \cite{dinkelbach1967nonlinear} and consider the following problem
\begin{align}
&{}{G(\lambda)=\max\limits_{\boldsymbol{x}\in\mathcal{C}}(1+\frac{1}{\sigma^{2}}(\sum\nolimits_{l=1}^{L}C_{l}x_{l}+\sum\nolimits_{l=2}^{L}\sum\nolimits_{m=1}^{l-1}C_{lm}x_{l}}\nonumber\\
&{}{x_{m}))-\lambda(1+\frac{1}{\sigma^{2}_{e}}(\sum\nolimits_{l=1}^{L}D_{l}x_{l}+\sum\nolimits_{l=2}^{L}\sum\nolimits_{m=1}^{l-1}D_{lm}x_{l}x_{m})),}\label{4-22}
\end{align}
where $\mathcal{C}$ denotes the feasible set of $\boldsymbol{x}$ satisfying constraint (\ref{4-19b}). According to \cite{dinkelbach1967nonlinear},
solving (\ref{4-22}) is equivalent to obtaining the root of $G(\lambda)$, and Dinkelbach method can obtain the root. After introduced the parameter $\lambda$,  (\ref{4-19a}) can be transformed as the formula in (\ref{4-22}).

To handle the non-convex constraint in (\ref{4-8d}), we introduce new variable $z_{lm}=x_{l}x_{m}$. {{}{}Owing to $x_{l}\in\{0,1\}$, constraint $z_{lm}=x_{l}x_{m}$ is equivalent to
\begin{eqnarray}
z_{lm}\geq x_{l}+x_{m}-1,
0\leq z_{lm}\leq 1,
z_{lm}\leq x_{l},
z_{lm}\leq x_{m}.\label{4-23}
\end{eqnarray}}
According to (\ref{4-22}) and (\ref{4-23}), problem (\ref{4-19}) is rewritten as
\begin{subequations}
\begin{align}
{}{\max\limits_{\boldsymbol{x},\boldsymbol{z}}}~&{}{(1+\frac{1}{\sigma^{2}}(\sum\nolimits_{l=1}^{L}C_{l}x_{l}+\sum\nolimits_{l=2}^{L}\sum\nolimits_{m=1}^{l-1}C_{lm}z_{lm}))}\nonumber\\
&{}{-\lambda(1+\frac{1}{\sigma_{e}^{2}}(\sum\nolimits_{l=1}^{L}D_{l}x_{l}+\sum\nolimits_{l=2}^{L}\sum\nolimits_{m=1}^{l-1}D_{lm}z_{lm})),}\label{4-27a}\\
{}{\mbox{s.t.}}~
&{}{\text{(\ref{4-8d})}},~{}{\text{(\ref{4-23})}},\label{4-27e}
\end{align}\label{4-27}%
\end{subequations}
where $\boldsymbol{z}=[z_{21},z_{31},\cdots,\cdots,z_{L(L-1)}]^{T}$.
Since constraint (\ref{4-8d}) is non-convex, handling problem (\ref{4-27}) is still difficult. To deal with this problem, we relax (\ref{4-8d}) with $x_{l}\in[0,1]$, then, the constraints in (\ref{4-27}) are convex. Problem (\ref{4-27}) can be rewritten as
\begin{subequations}
\begin{align}
{}{\max\limits_{\boldsymbol{x},\boldsymbol{z}}}~&{}{(1+\frac{1}{\sigma^{2}}(\sum\nolimits_{l=1}^{L}C_{l}x_{l}+\sum\nolimits_{l=2}^{L}\sum\nolimits_{m=1}^{l-1}C_{lm}z_{lm}))}\nonumber\\
&{}{-\lambda(1+\frac{1}{\sigma^{2}_{e}}(\sum\nolimits_{l=1}^{L}D_{l}x_{l}+\sum\nolimits_{l=2}^{L}\sum\nolimits_{m=1}^{l-1}D_{lm}z_{lm}))},\label{4-28a}\\
{}{\mbox{s.t.}}~
&{}{\text{(\ref{4-23})},}\label{4-28e}\\
&{}{x_{l}\in[0,1].}\label{4-28f}
\end{align}\label{4-28}%
\end{subequations}
For problem (\ref{4-28}) with relaxed constraints, the optimal solution can be obtained by the dual method in \cite{boyd2004convex}. The integer solution is obtained by the dual method and it guarantees both optimality and feasibility of the original problem. To obtain the optimal solution of problem (\ref{4-28}), we give the following theorem.
\begin{theorem}\label{The3}
For problem (\ref{4-28}), variables $\boldsymbol{x}$ and $z_{lm}$ are respectively denoted as
\begin{eqnarray}
{}{x_{l}=\left\{
\begin{aligned}
1 &~& S_{l}> 0 \\
0 &~& S_{l}\leq 0,\label{4-26} 
\end{aligned}
\right.}~~~~
{}{z_{lm}=\left\{
\begin{aligned}
1 &~& S_{lm}< 0 \\
0 &~& S_{lm}\geq 0,\label{4-27m} 
\end{aligned}
\right.}
\end{eqnarray}
where $S_{l}$ is given at the top of next page
\newcounter{mytempeqncnt}
\begin{figure*}[!t]
\normalsize
\setcounter{mytempeqncnt}{\value{equation}}
\begin{eqnarray}
{}{ S_{l}=\left\{
\begin{aligned}
&\sum\nolimits_{m=2}^{L}(\lambda_{ml}^{1}+\lambda_{ml}^{2}+\lambda_{ml}^{3})+(\frac{1}{\sigma^{2}_{e}}\lambda-\frac{1}{\sigma^{2}})D_{l},&~l=1,\\
&\sum\nolimits_{m=1}^{l-1}(\lambda_{lm}^{1}+\lambda_{lm}^{2})+\sum\nolimits_{m=l+1}^{L}(\lambda_{ml}^{3}+\lambda_{ml}^{1})+(\frac{1}{\sigma^{2}_{e}}\lambda-\frac{1}{\sigma^{2}})D_{l}, &~2\leq l\leq L-1, \\
&\sum\nolimits_{m=1}^{L-1}(\lambda_{lm}^{1}+\lambda_{lm}^{2})+(\frac{1}{\sigma^{2}_{e}}\lambda-\frac{1}{\sigma^{2}})D_{l}, &~l=L\\
\end{aligned}\label{4-28m}
\right.}
\end{eqnarray}
\setcounter{equation}{\value{mytempeqncnt}}
\hrulefill
\vspace*{4pt}
\end{figure*}
and
\setcounter{equation}{25}
\begin{eqnarray}
{}{\bar{S}_{lm}=(\lambda_{lm}^{1}+\lambda_{lm}^{2}+\lambda_{lm}^{3})+(\frac{1}{\sigma^{2}}-\frac{1}{\sigma^{2}_{e}}\lambda)D_{lm},}\label{4-29}
\end{eqnarray}
where $\{\lambda_{lm}^{1},\lambda_{lm}^{2},\lambda_{lm}^{3},\lambda\}$ are the Lagrange multipliers associated with corresponding constraints of problem (\ref{4-28}).

The proof is given in \textbf{Appendix}~\ref{appC}.
\end{theorem}
\textbf{Theorem}~\ref{The3} states that the $l$th IRS has a positive coefficient $S_{l}$ should be on.  According to the  expression of $S_{l}$ in (\ref{4-19}), the negative term, $-\frac{1}{\sigma^{2}} A D_{l}$, is the effect of introducing additional interference from eavesdropper when the $l$th IRS is on. At the same time, the remaining part denotes the benefit of increasing the user's rate by keeping IRS $l$ in operation. When $S_{l}>0$, the benefit of increasing the user's rate is larger than the effect of introducing additional interference from eavesdropper, which means that the secrecy rate can be improved when the $l$th IRS is on. 
The values of $\{\lambda_{lm}^{1},\lambda_{lm}^{2},\lambda_{lm}^{3},\lambda\}$ are updated by the subgradient method \cite{boyd2004convex}, they are denoted as
\begin{eqnarray}
&{}{\lambda_{lm}^{1}=[\lambda_{lm}^{1}-\beta(z_{lm}-x_{l}-x_{m}+1)]^{+},}&\label{4-32}\\
&{}{\lambda_{lm}^{2}=[\lambda_{lm}^{2}-\beta(z_{lm}-x_{l})]^{+},}&\label{4-33}\\
&{}{\lambda_{lm}^{3}=[\lambda_{lm}^{3}-\beta(z_{lm}-x_{m})]^{+},}&\label{4-34}
\end{eqnarray}
where $\beta>0$ is a step-size sequence. \par
By iteratively optimizing $(\boldsymbol{x},\boldsymbol{z})$ and $\{\lambda_{lm}^{1},\lambda_{lm}^{2},\lambda_{lm}^{3},\lambda\}$, the optimal
$\boldsymbol{x}$ is obtained. The dual method for solving problem (28) and the Dinkelbach method to update parameter $\lambda$ are given in \textbf{Algorithm}~\ref{algo-2}. It is not difficult to find that the optimal $x_{l}$
is either 0 or 1 according to (\ref{4-26}), even though $x_{l}$ is relaxed as (\ref{4-28f}). 
Using the Dinkelbach method, we can obtain the root of $G(\lambda)$ = 0, which indicates that the optimal solution of the secrecy rate optimization problem in (\ref{4-19}) is obtained. 
\begin{algorithm}[!htbp]\small
\caption{Proposed Langrange Dual Algorithm for Problem (\ref{4-19})}
   \begin{algorithmic}[1]
   \REQUIRE  $t=0$,  
$\lambda^{0}$ and set the accuracy $\epsilon$.\\ 
   \STATE \textbf{repeat}
.     \STATE Initialize $\{\lambda_{lm}^{1},\lambda_{lm}^{2},\lambda_{lm}^{3},\lambda\}^{0}$.\\
\STATE \textbf{repeat}
    \STATE Update the IRS on-off vector $\boldsymbol{x}$ according to (\ref{4-29}).\\
   \STATE Update dual variables $\{\lambda_{lm}^{1},\lambda_{lm}^{2},\lambda_{lm}^{3},\lambda\}^{0}$ based on (\ref{4-32})-(33).\\
    \STATE $t=t+1$.\\
     \STATE \textbf{until} the objective value converges.\\
     \STATE Denote the objective value (\ref{4-28a}) by $G(\lambda)$.\\
     \STATE Update $\lambda$ based on $G(\lambda)=0$.\\
    \STATE $t_{1}=t_{1}+1$.\\
     \STATE \textbf{until} $G(\lambda)<\epsilon$.\\
     \ENSURE $\boldsymbol{x}^{*}$.\\
   \end{algorithmic}\label{algo-2}
 \end{algorithm}

Then, given $\boldsymbol{x}$ and $\{\boldsymbol{w}\}$, problem (\ref{4-8}) can be simplified as
{{}{}\begin{subequations}
\begin{align}
{}{\max\limits_{\boldsymbol{\theta}}}~&{}{\log_{2}(1+\frac{1}{\sigma^{2}}\|\sum\nolimits_{l=1}^{L}x_{l}\boldsymbol{h}_{l}^{H}\boldsymbol{\Theta}_{l}\boldsymbol{G}_{l}\boldsymbol{w}\|^{2})}\nonumber\\
&{}{-\log_{2}(1+\frac{1}{\sigma_{e}^{2}}\|\sum\nolimits_{l=1}^{L}x_{l}\boldsymbol{g}_{l}^{H}\boldsymbol{\Theta}_{l}\boldsymbol{G}_{l}\boldsymbol{w}\|^{2}),}\label{4-33a}\\
{}{\mbox{s.t.}}~
&{}{\text{(\ref{4-8c})}.}\label{4-33b}
\end{align}\label{4-33m}%
\end{subequations}}%

The problem (\ref{4-33m}) can be solved efficiently by the manifold optimization (MO) algorithm as \cite{absil2009optimization}. Details are omitted for
simplicity.

\subsection{Complexity Analysis}
The total computational complexity of each iteration of the proposed AO algorithm is $\mathcal{O}(\max\{N_{t}^{3.5}\log(1/\epsilon),$ $ TL^{2}, LN_{r}\})$. The proposed algorithm based on BCD has lower computational complexity than the PGD algorithm [17] of $\mathcal{O}(\max\{(N_{t}+1)^{3.5}\log(1/\epsilon), MN_{t}^{2}\})$ and the SDP [3] algorithm of $\mathcal{O}(N_{t}^{8}+MN_{t})$. 

\begin{table}[htbp]
  \centering
  \scriptsize
  \caption{Comparison of Algorithm Complexity}
  \label{tab:notations}
  \begin{tabular}{ll}
    \\[-2mm]
    \hline
    \hline\\[-2mm]
    {\bf \small Symbol}&\qquad {\bf\small Total Complexity}\\
    \hline
    \vspace{0.7mm}\\[-2mm]
    Proposed AO-based algorithm      &   $\mathcal{O}(\max\{N_{t}^{3.5}\log(1/\epsilon), TL^{2}, LN_{r}\})$\\
    \vspace{0.7mm}
    PGD-based algorithm~\cite{feng2019secure}          &  $\mathcal{O}((N_{t}+1)^{3.5}\log(1/\epsilon))$\\
     \vspace{0.7mm}
    SDP-based algorithm~[3]          &  $\mathcal{O}(N_{t}^{8}+MN_{t})$\\
    \hline
    \hline
  \end{tabular}
\end{table}\vspace{-20pt}

\section{Numerical Results}\label{IV}
As shown in Fig.~\ref{fig:1}, AP's coordinate is $(0,0,0)$ and three IRSs are
located at $(0, 20, 20)$~m, $(0, 40, 20)$~m, and $(0, 60, 20)$~m , respectively.  While the user and the eavesdropper are located at $(5,40,0)$ and $(5,60,0)$ in meters, respectively.
{{}{}The the AP-to-$l$th IRS, the $l$th IRS-to-user, and the $l$th IRS-to-eavesdropper mmWave channels are respectively expressed as \cite{xiao2017millimeter}
\begin{eqnarray}
{}{\boldsymbol{G}_{l}=\sqrt{1/\beta_{l}L_{1}}\sum_{l=0}^{L_{1}-1}\alpha_{l}\boldsymbol{a}_{T}(N_{T},\theta_{l})\boldsymbol{a}_{R}^{T}(N_{r},\varphi_{l},\phi_{l}),}
\end{eqnarray}
\begin{eqnarray}
{}{\boldsymbol{h}_{l}=\sqrt{1/\hat{\beta}_{l}L_{2}}\sum_{l=0}^{L_{2}-1}\hat{\alpha}_{l}\boldsymbol{a}_{R}(N_{r},\vartheta_{l}),}
\end{eqnarray}
and
\begin{eqnarray}
{}{\boldsymbol{g}_{l}=\sqrt{1/\bar{\beta}_{l}L_{3}}\sum_{l=0}^{L_{3}-1}\bar{\alpha}_{l}\boldsymbol{a}_{R}(N_{r},\varepsilon_{l}),}
\end{eqnarray}
where $\beta_{l}$, $\hat{\beta}_{l}$, and $\bar{\beta}_{l}$ denote the large-scale fading coefficients. They are generated by $\zeta-10c\log_{10}(d)$, where $d$ is the signal propagation distance. $c$ is the path loss exponent, $\zeta=-61.4$~dB, $\alpha_{l}$, $\hat{\alpha}_{l}$, and $\bar{\alpha}_{l}$ denote the small-scale fading coefficients which follow $\mathcal{CN}(0,1)$ [12]. $\boldsymbol{a}_{T}(\cdot)$ and $\boldsymbol{a}_{R}(\cdot)$ represent array steering vectors at the AP and the IRS, respectively.}
We set $N_{t}=16$, $N_{r}=16$, and $\sigma^{2}=\sigma^{2}_{e}=-110$~dBm. We compare the proposed scheme with the single IRS-aided scheme, where the IRS in the single IRS-aided scheme locates at $(0,60,20)$ in meters, and the number of reflecting elements of the IRS is set as the total number of reflecting elements for all IRSs in multiple IRSs-aided scheme.

\begin{figure*}[htbp]
\centering
\subfigure{
\begin{minipage}[t]{0.3\linewidth}
\centering
\includegraphics[scale=0.27]{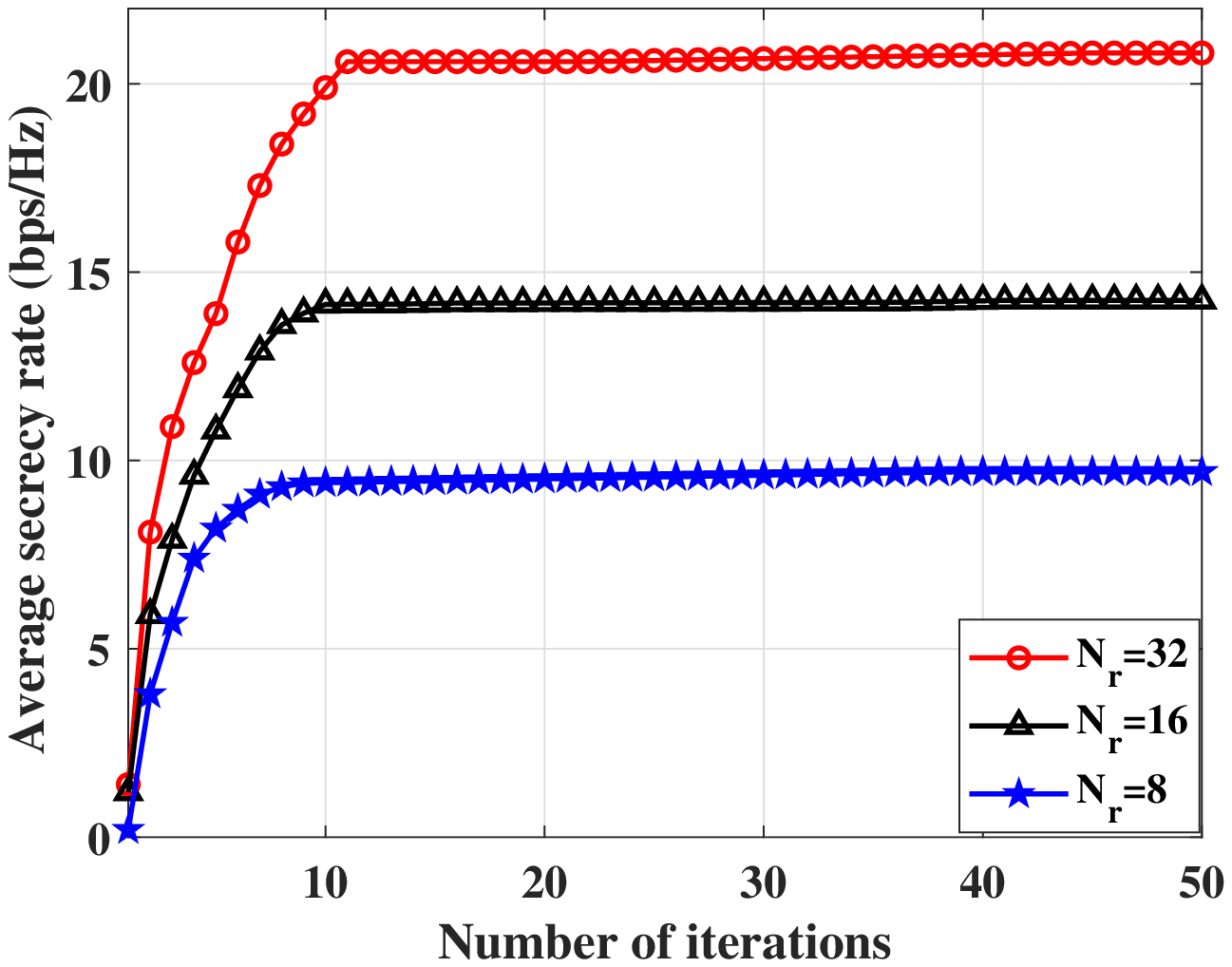}
\caption{Convergence of Algorithm}
\label{fig:2a}
\end{minipage}%
}%
\subfigure{
\begin{minipage}[t]{0.25\linewidth}
\centering
\includegraphics[scale=0.27]{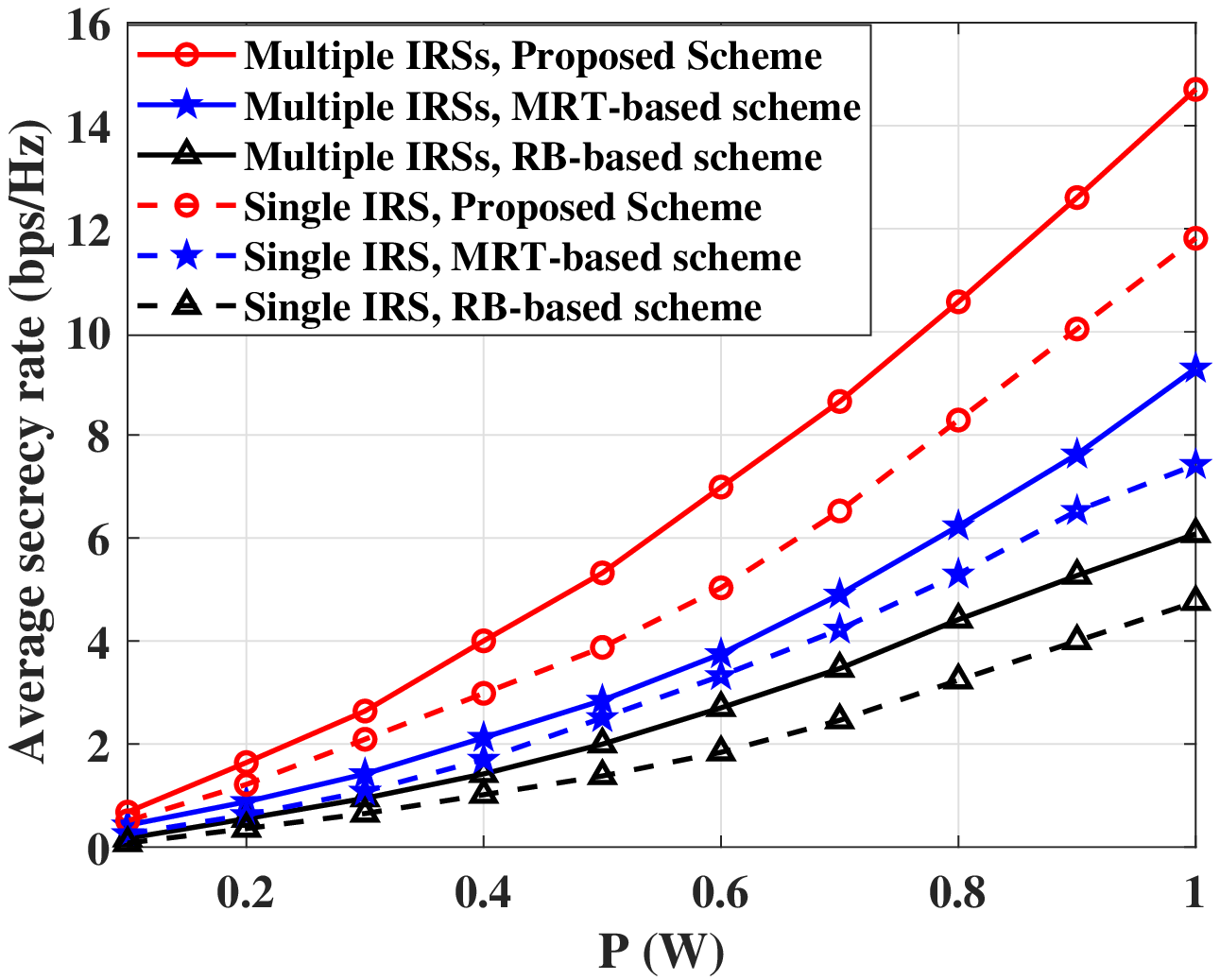}
\caption{ASR versus $P$.}
\label{fig:2}
\end{minipage}%
}%
\subfigure{
\begin{minipage}[t]{0.25\linewidth}
\centering
\includegraphics[scale=0.27]{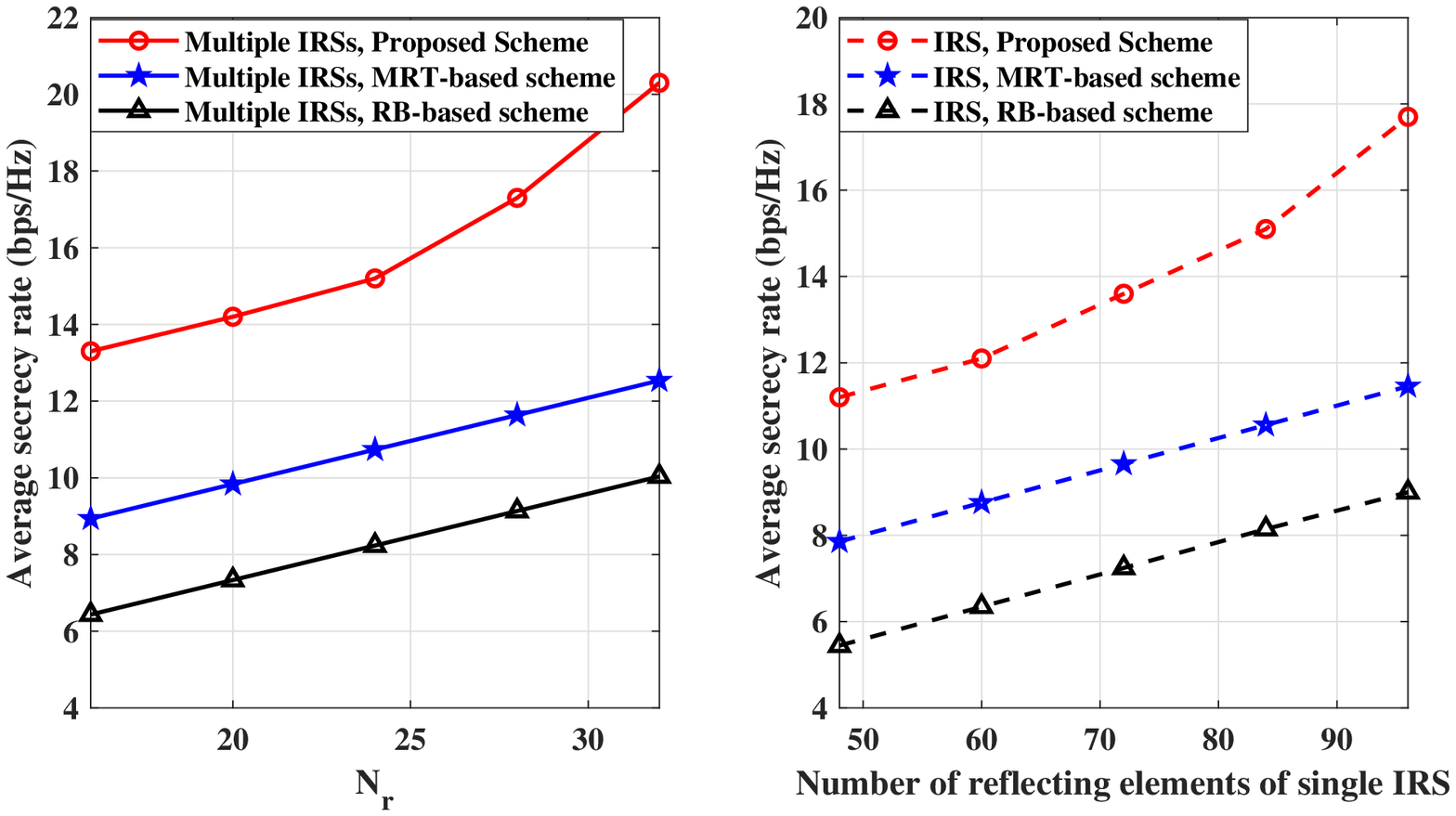}
\caption{ASR versus $M$.}
\label{fig:4}
\end{minipage}
}%
\centering
\end{figure*}

{{}{}In Fig.~\ref{fig:2a}, we study the convergence behavior of the proposed algorithm versus different numbers of reflecting elements. It can be observed that the secrecy rate first increases monotonically with the number of iterations. Then, the proposed algorithm converges rapidly. In general, a few iterations are sufficient for the proposed algorithm to achieve a high secrecy rate. This shows the complexity of the proposed algorithm is low.}

Fig.~\ref{fig:2} depicts the average secrecy rate (ASR) versus the power under different beamforming schemes. The proposed algorithm achieves the best performance. From Fig.~\ref{fig:2}, the mmWave system with multiple IRSs can increase up to 20\% secrecy rate compared with the mmWave system with a single IRS. This is due to the benefits of multiple deployment. Multiple IRSs are spatially distributed in multiple IRSs-aided scheme, which can provide more than one path of the received signal compared with one central IRS. In addition, the ASR of the proposed AO-based algorithm significantly increases, but those of maximum ratio transmission (MRT)-based and random beamforming (RB)-based schemes increase slowly. This is the fact that the MRT-based and RB-based scheme aims to maximize the achievable rate of the user while ignoring the eavesdropper, which results in significant information leakage.

Fig.~\ref{fig:4} plots the secrecy rate versus the number of reflecting elements. As shown in Fig.~\ref{fig:4}, the secrecy rate monotonically increases with the number of reflecting elements. This is because a large number of reflecting elements can lead to a high signal gain and suppress the eavesdropper, which results in a high secrecy rate of the system. In addition, the mmWave system with multiple IRSs can perform better than that with one IRS. It shows that multiple IRSs-aided scheme is more efficient in improving secrecy rate of mmWave system. 

\section{Conclusion and Future Works}
The secrecy rate maximization problem for mmWave communications with multiple IRSs was investigated. The IRS phase shifts, transmit beamforming, and IRS on-off status were jointly optimized to maximize the secrecy rate under the transmit power constraint and unit-modulus constraints. To solve this non-convex problem, we have proposed an AO-based algorithm. Numerical results show that the proposed AO-based algorithm outperforms the traditional schemes in terms of secrecy rate.

\appendices
\section{The proof of \textbf{theorem}\ref{The3}}\label{appC}
The Lagrange function of (\ref{4-28}) with relaxed constraints is expressed as
\begin{align}
&{}{L(\boldsymbol{x},\boldsymbol{z},\lambda_{lm}^{1},\lambda_{lm}^{2},\lambda_{lm}^{3})=\big(1+\frac{1}{\sigma^{2}}(\sum\nolimits_{l=1}^{L}C_{l}x_{l}+\sum\nolimits_{l=2}^{L}\sum\nolimits_{m=1}^{l-1}}\nonumber\\
&{}{C_{lm}z_{lm})\big)-\lambda(1+\frac{1}{\sigma_{e}^{2}}(\sum\nolimits_{l=1}^{L}D_{l}x_{l}+\sum\nolimits_{l=2}^{L}\sum\nolimits_{m=1}^{l-1}D_{lm}z_{lm}))}\nonumber\\
&{}{+\sum\nolimits_{l=2}^{L}\sum\nolimits_{m=1}^{l-1}(\lambda_{lm}^{1}(z_{lm}-x_{l}-x_{m}+1)+\lambda_{lm}^{2}(z_{lm}-x_{l})}\nonumber\\
&{}{+\lambda_{lm}^{3}(z_{lm}-x_{m})).}
\end{align}
To maximize the objective function in (\ref{4-28a}), let $\frac{\partial L(\boldsymbol{x},\boldsymbol{z},\lambda_{lm}^{1},\lambda_{lm}^{2},\lambda_{lm}^{3})}{\partial x_{l}}=0$ and $\frac{\partial L(\boldsymbol{x},\boldsymbol{z},\lambda_{lm}^{1},\lambda_{lm}^{2},\lambda_{lm}^{3})}{\partial z_{lm}}=0$, thus, when $l=1$, we have
\begin{align}
{}{\frac{1}{\sigma^{2}}C_{l}-\frac{1}{\sigma_{e}^{2}}\lambda D_{l}-\sum\nolimits_{m=2}^{L}(\lambda_{ml}^{1}+\lambda_{ml}^{2}+\lambda_{ml}^{3})=0.}\label{4-35}
\end{align}
When $2\leq l\leq N-1$, we have
\begin{align}
&{}{\frac{1}{\sigma^{2}}C_{l}-\frac{1}{\sigma_{e}^{2}}\lambda D_{l}-\sum\nolimits_{m=1}^{l-1}(\lambda_{lm}^{1}+\lambda_{lm}^{2})-\sum\nolimits_{m=l+1}^{L}(\lambda_{ml}^{3}}\nonumber\\
&{}{+\lambda_{ml}^{1})=0.}\label{4-36}
\end{align}
When $l=N$, we have
\begin{align}
&{}{\frac{1}{\sigma^{2}}C_{l}-\frac{1}{\sigma_{e}^{2}}\lambda D_{l}-\sum\nolimits_{m=1}^{L-1}(\lambda_{lm}^{1}+\lambda_{lm}^{2})=0.}\label{4-37}
\end{align}
Based on $\frac{\partial L(\boldsymbol{x},\boldsymbol{z},\lambda_{lm}^{1},\lambda_{lm}^{2},\lambda_{lm}^{3})}{\partial z_{lm}}=0$, the relationship between $C_{lm}$ and $D_{lm}$ is given by
\begin{align}
{}{\frac{1}{\sigma^{2}}C_{lm}-\frac{1}{\sigma_{e}^{2}}\lambda D_{lm}+(\lambda_{lm}^{1}+\lambda_{lm}^{2}+\lambda_{lm}^{3})=0.}\label{4-38}
\end{align}
To maximize the objective function in (\ref{4-19a}), it is equivalent to maximize the $(18)/(19)$.
When the $l$th IRS is on, we hope the user gain generated by IRS is larger than the eavesdropper gain, i.e.,
\begin{align}
{}{C_{l}>D_{l}, C_{lm}>D_{lm}.}\label{4-39m1}
\end{align}
According to (\ref{4-35})-(\ref{4-38}) and (\ref{4-39m1}), we have the following inequalities in (\ref{4-28m}) and (\ref{4-29}), when $S_{l}>0$, $C_{l}>D_{l}$. So, $x_{l}$ is set as $1$. The \textbf{Theorem}~\ref{The3} is proved. \vspace{-12pt}

\ifCLASSOPTIONcaptionsoff
  \newpage
\fi



\end{document}